\documentstyle[11pt]{article}
\newcommand{\be}{\begin{equation}}
\newcommand{\ee}{\end{equation}}
\newcommand{\ba}{\begin{eqnarray}}
\newcommand{\ea}{\end{eqnarray}}
\newcommand{\bann}{\begin{eqnarray*}}
\newcommand{\eann}{\end{eqnarray*}}
\newcommand{\uu}{\underline}

\begin{document}
\hbadness=10000
\setcounter{page}{1}

\title{Bose-condensation through resonance decay}

\author{U. Ornik$^1$\thanks{E. Mail: ORNIK@TPRI6B.GSI.DE},
M. Pl\"umer$^2$\thanks{
E. Mail:PLUEMER\_M@VAX.HRZ.UNI-MARBURG.DE} and
D. Strottman$^3$ \thanks{E. Mail: DDS@T2.LANL.GOV}}

\date{$^1$  Theory Group, Gesellschaft f\"ur Schwerionenforschung (GSI),
Darmstadt, F.R. Germany\\
$^2$Physics Department, University of Marburg, FRG\\
$^3$ Theoretical Division, Los Alamos National Laboratory, Los Alamos, USA
87545
\\
\today}

\maketitle

\begin{abstract}
We show that a system described by an equation of state which
contains a high number of degrees of freedom (resonances) can
create a considerable amount of superfluid (condensed) pions through
the decay of short-lived resonances, if baryon number
and entropy are large and the dense matter
decouples from chemical equilibrium earlier than
from thermal equilibrium.
The system cools down faster in the
presence of a condensate, an effect that may partially
compensate the enhancement
of the lifetime expected in the case of quark-gluon-plasma
formation.

\end{abstract}

\newpage

The investigation of hot hadronic matter through heavy ion
collisions has shown that an understanding of medium effects and
final state interactions is of great importance for the interpretation
of the particle spectra in order to distinguish new physical
phenomena like the formation of quark-gluon plasma (QGP) from
background phenomena caused by ``conventional'' physical effects.
We refer here, e.g., to the strangeness enhancement, the $J/\Psi$
suppression and the soft-pion puzzle, which can in principle be
explained either by the formation of a QGP or from purely hadronic
origin
(for a recent review, see ref. \cite{QM91}).

An interesting explanation for the
soft pion puzzle was proposed in \cite{leutwyler}.
The authors argued that
the baryonic and mesonic
resonances may decouple from chemical equilibrium but still
remain in thermal equilibrium. Since the resonance production rates fall
drastically after their decoupling from chemical equilibrium, the
remaining short-lived resonances would then rapidly
decay into pions.
As a consequence, the pions may acquire a non-zero chemical potential
$\mu_\pi$
which leads to a softer $p_\perp$-spectrum.

A different possible explanation for the absence of
local chemical equilibrium of pions in the final (decoupling)
stage, which should not be confused with the scenario discussed
in  ref. \cite{leutwyler}, was proposed in refs.\cite{sean,bertsch}
where it is assumed that the pions are initially created out of local
chemical (and even thermal!) equi\-lib\-rium
and it turns out that the system
never reaches complete local chemical
equi\-lib\-rium\cite{sean,bertsch}.

In the present letter, we investigate the scenario
of ref. \cite{leutwyler}, where it is assumed that the
dense matter reaches local thermodynamic equilibrium in an early stage
of the expansion. The authors of
\cite{leutwyler} considered only the case that
the chemical potential stays below the pion mass. Moreover,
they did not attempt to determine the value of the pion chemical
potential in terms of the temperature and baryon number
density at the point of decoupling from local chemical equilibrium.
Below, we address the problem if and under what
conditions the decay of short-lived resonances can lead to the
formation of a pionic Bose condensate, and what happens to the
condensate
in the subsequent expansion until the system decouples from local
thermal equilibrium
and the final state particles are emitted.
To do so, we explicitly
consider the transition from a resonance gas to a gas of stable
and long-lived particles for a hadronic equation of state which
contains the known resonances up to masses of $2$ GeV, imposing
constraints from the conservation of energy-momentum, baryon
number and strangeness.
In particular, we shall
show that for a system rich in entropy and baryon number
$\mu_\pi$ may reach the pion mass and, consequently, the pions may
form a Bose condensate in the central region
in relativistic
heavy ion collisions.

For a hydrodynamically expanding
system, the formation of such a condensate implies the presence of
a superfluid component. The idea that the
hadronic matter might be superfluid was put forward a long time ago in a
somewhat different context \cite{weiner_old}.
This could explain in a natural way the presence of a coherent
component in multiparticle production. The implications for pion
interferometry measurements will be discussed in a separate paper
\cite{becon}.
We note that
although the creation of a remarkable amount of condensed pions is
interesting by itself, these effects could also be used to answer the
question of whether pions and resonances are in chemical
equilibrium before they decouple from thermal equilibrium.
Because the final stage in the hadronic phase can have a
significant influence on the particle spectrum, the question of
whether the freeze-out occurs in chemical and thermal equilibrium,
as assumed in \cite{ornik_soft,heinz} or only in thermal equilibrium
\cite{leutwyler} is of great importance.

In order to investigate the conditions necessary to form a
condensate, let us first consider a system in local thermodynamic
(i.e., thermal {\it and} chemical) equilibrium. The medium is then
completely described in terms of its
equation of state (EOS) which can be expressed in the form

\be
\varepsilon_{fl}=f(T_{fl},\mu_{fl}^k) \label{eq:EOS_def}
\ee

where $\varepsilon_{fl}$ is the energy-density of the fluid,
$T_{fl}$ the temperature, and the quantities $\mu_{fl}^k$
are the chemical potentials related to conserved charges
$Q^{k}$ (such as baryon number
$B$ and strangeness $S$).
The index $fl$
refers to quantities which describe the fluid.

During a heavy ion collision we expect under certain conditions the
formation of a system of hot hadronic matter or a QGP described by
eq. (\ref{eq:EOS_def}) which is in chemical and thermal equilibrium.
The system then undergoes a phase of hydrodynamic expansion until it
has become so dilute that the interactions are no longer strong enough
to maintain local equilibrium. This final stage,
when the collective behaviour terminates
and particles decouple from the dense matter,
is referred to as freeze-out.

The scenario of ref.\cite{leutwyler} differs from most of the
other hydrodynamic descriptions of heavy ion collisions in so
far as in \cite{leutwyler}
the dense matter is assumed to go out of chemical
equilibrium but remain in thermal equilibrium for some time.
Or, to put it differently, there will be a ``chemical freeze-out''
independent of and prior to the ``thermal freeze-out''.
As was mentioned above, the transition out of chemical equilibrium
leads to the decay of short-lived resonances and the appearance
of a non-zero chemical potential of the pions.

Below, we shall assume that the chemical freeze-out, just like the
thermal freeze-out, occurs ``instantaneously'' (i.e., on a
three-dimensional hypersurface characterized by a condition
such as $T(x,t)=const.$). This seems a reasonable assumption since
the relevant time scales are
the decay times of the short-lived resonances
which are on the order of $\sim 1$ fm/c.
To illustrate the
rapidness of such a transition, we consider the simplified
case of a system
of pions and rho-mesons in local thermal equilibrium
that undergoes a longitudinal scaling
\cite{Bj} expansion. In addition to the decay
$\rho \rightarrow \pi \pi$ we also take into account
the inverse process $\pi\pi
\rightarrow \rho + X$. The equations which describe the time evolution
of the system are
\ba
\frac{d\varepsilon}{d\tau} & = & - \frac{\varepsilon+P}{\tau}
\label{eq:euler}\\
\frac{dn_\pi}{d\tau} & = & - \frac{n_\pi}{\tau} \ + \
2 \ \frac{n_\rho}{\tau_\rho} \ - \ 2
\ \langle \sigma v \rangle n_\pi^2 \label{eq:npi}\\
\frac{dn_\rho}{d\tau} & = & - \frac{n_\rho}{\tau} \ - \
 \frac{n_\rho}{\tau_\rho} \ + \  \langle \sigma v \rangle n_\pi^2
\label{eq:nrho}
\ea
where $\varepsilon = \varepsilon_\rho + \varepsilon_\pi$ and
$P=P_\rho +P_\pi$ are the energy density and pressure, $n_i$
$(i=\rho,\pi)$ are the number densities and $\tau \equiv \sqrt{t^2-z^2}$
is the longitudinal propertime coordinate. $\tau_\rho$ is the lifetime of
the $\rho$, and $\langle \sigma v \rangle$ is the thermal average of the
cross section for the process $\pi \pi \rightarrow \rho +X$.
We have solved eqs.(\ref{eq:euler}-
\ref{eq:nrho}) numerically
under the assumption that at $\tau=\tau_0$ the $\pi$'s and $\rho$'s
are in chemical equilbrium at a temperature $T_0$,
and that the system
remains in local thermal equilibrium during the expansion
(note that the eqs.(\ref{eq:npi},\ref{eq:nrho}) imply that
$(n_\pi +2 n_\rho)\tau = const.$). Fig. 1 shows the
time evolution of the pion chemical potential $\mu_\pi$ for
$\tau_0=1$ fm/c and $T_0=200$ MeV. It turns out that for reasonable values
$\langle \sigma v \rangle\ \leq \ 100$ mb, contributions of the inverse
process are negligible.
It can be seen that $\mu_\pi$
approaches its final value very fast, after less than $1$ fm/c.
For larger values of $\tau_0$, the transition is even faster.
The figure also shows that the decay of rho-mesons alone is not
sufficient to drive the pions into the condensate (for that, one
needs to take into account the decay contributions of all the other
short-lived resonances as well).

We now return to the description of the expanding hot and dense
matter at the point of chemical freeze-out.
It will be assumed that the fluid freezes out into ``stable''
$\pi$'s , $K$'s,
nucleons, $\Lambda$'s and the long living resonances
$\omega$'s and $\eta$'s. The term ``stable'' here means
that the decay of the particle or resonance occurs only after the
complete (thermal and chemical) freeze-out of the system.
In particular, the contribution of $\omega$'s and $\eta$'s to the
pion chemical potential is zero. Nevertheless, their later decay
leads to a non-thermal component of the pion spectrum which
is determined by the decay kinematics \cite{ornik_soft} .

The system of chemically frozen out particles can be described
by a chemical decoupling temperature $T^{ch.f.}$, chemical
potentials for baryons and strange
particles, $\mu_B$ and $\mu_S$,
and a pion chemical potential, $\mu_\pi$, which describes the
overpopulation of pions due to the decoupling of
resonances. The chemical potentials are determined by the
requirement that the energy density $\varepsilon_{fl}$, the
baryonic density $b_{fl}$ and the strangeness density $s_{fl}=0$
of the fluid are equal to those of the system after chemical freeze-out.
We obtain the following system of equations:

\ba
\varepsilon_{fl} &=&\varepsilon^{therm}_\pi(\mu_\pi,T^{ch.f.})
+m_\pi n_\pi^{con}
\Theta(m_\pi-\mu_\pi)\nonumber\\
&& + \sum_{i=K,N,\Lambda,\omega,\eta,...}
 \varepsilon^{therm}_i(\mu_S,\mu_B,T^{ch.f.})
\label{eq:coupled_eq1}\\
b_{fl}&=&\sum_k B_k \
n^{therm}_k(\mu_S,\mu_B,\mu_\pi,T^{ch.f.})\label{eq:coupled_eq2}\\
s_{fl}&=&\sum_k S_k \
n^{therm}_k(\mu_S,\mu_B,\mu_\pi,T^{ch.f.})\ =\ 0 \label{eq:coupled_eq3}
\ea

where

\ba
\varepsilon^{therm}_i(\mu_S,\mu_B,\mu_\pi,T)
&=&\frac{g_i}{2\pi^2}\int p^2dp\frac{E_i}
                     {exp\left(\frac{E_i-\tilde\mu_i}{T}\right)\pm1}
\label{eq:etherm}\\
n^{therm}_i(\mu_S,\mu_B,\mu_\pi,T)
&=&\frac{g_i}{2\pi^2}\int p^2dp\frac{1}
                     {exp\left(\frac{E_i-\tilde\mu_i}{T}\right)\pm1}
\label{eq:ntherm}
\ea

are the thermal parts of the energy density and the number density
of particle species $i$,
and where we have used the notation

\be
\tilde\mu_i=\sum_k Q_i^k\mu^k
\label{eq:potential}
\ee

for the chemical potentials.
Note that on the r.h.s. of eq. (\ref{eq:potential})
the charges $Q^k$ include the pion number $N_\pi$ which is a conserved
quantity after chemical freeze-out.

In eq. (\ref{eq:coupled_eq1}) we have introduced a term $n_\pi^{con}$
that describes the condensed component of pions  which will
appear if the chemical equilibrium for pions reaches the pion mass.
In this case an overpopulation of pion states occurs and the
remaining pions will be forced into the ground state,
thereby forming a Bose condensate.
The fluid then consists of a thermal component and a
superfluid component.
The latter moves with the rapidity of the
fluid, whereas the thermal pions have on top of the fluid rapidity a
thermal distribution.
The resultant single inclusive
distribution of pions emitted from a hydrodynamically
expanding source

\ba
\frac{1}{p_\perp}\frac{dn}{dydp_\perp} & = &
\frac{g_\pi}{(2\pi)^3}\int_\sigma
\frac{p^i_\mu d\sigma^\mu}{exp\left(\frac{p_\mu^i u^\mu-\tilde\mu_\pi}
{T_{fl}}\right) - 1}\nonumber\\
& &  + \int_\sigma n_\pi^{con} u_\mu d\sigma^\mu
\delta(y-\tanh\left(u_{||}/u_0\right) )
\delta^2(\vec p_\perp-m_\pi \vec u_\perp)\label{eq:hydro}
\ea
where $y$ and $p_\perp$ are the rapidity and transverse momentum of
the emitted pion, $u^\mu$ is the
four-velocity field and $\sigma$ is the freeze-out hypersurface.
The first term on the r.h.s. of (\ref{eq:hydro}) is
the thermal contribution given
by the relativistic invariant Cooper-Frye formula \cite{cooper},
and the second term
is the condensate contribution.

The coupled system of nonlinear of equations
(\ref{eq:coupled_eq1})-(\ref{eq:coupled_eq3}) have to be solved in
order to determine from the fluid variables $\varepsilon_{fl}$,
$n_{fl}$ and $s_{fl}=0$ the quantities $\mu_S$, $\mu_B$ and $\mu_\pi$
 (for $\mu_\pi < m_\pi$), or $\mu_S$, $\mu_B$ and $n_\pi^{con}$ (for
$\mu_\pi =m_\pi$)
which describe the system after chemical freeze-out.

Let us now investigate
under what circumstances the system
will develope a superfluid component. To this end,
we apply a hadronic
EOS which contains
all the important resonances and a treatment
of compression effects, thus
retaining the essential features of nuclear matter near
the ground state \cite{dan}. In addition to the work presented in
\cite{dan} we take into account the effects of finite baryonic and
mesonic masses,
use Fermi statistics for the baryons and include mesons up
to masses of 2 GeV.

Fig. 2 shows the thermal and the condensate
component of pions at chemical freeze-out, and the chemical
potentials $\mu_\pi$, $\mu_b$ and $\mu_s$, as functions of the
temperature $T^{ch.f.}$ that characterizes the decoupling from
chemical equilibrium, for three different values of the
baryon number density $b$.
It can be seen
how the baryonic and strange chemical potentials,
$\mu_b$ and
$\mu_s$, decrease with increasing temperature $T^{ch.f.}$.
Clearly, the strange
chemical potential induced by the finite baryon number
never execeeds 150 MeV.
As the amount of heavier resonances increases with temperature,
the pion chemical potential $\mu_\pi$ grows with $T^{ch.f.}$.
It becomes equal to the pion mass in the
temperature region 170-200 MeV; the value of $T^{ch.f.}$ where
$\mu_\pi=m_\pi$ depends on the baryonic density.
This behaviour is also reflected in the dependence of the
thermal and condensate densities of pions on the temperature
and baryon number density at chemical freeze-out.
It can be seen that at sufficiently high baryonic densities
and temperatures
the condensate contribution to the chemically decoupled pions
becomes significantly large.
It is noteworthy that
the condensation effect is enhanced if the density of the strange
particles has not yet reached its equilibrium value. In this case
the remaining energy is distributed among the non strange resonances,
and this leads to an enhanced production of
particles in the condensate.

Finally, we need to discuss the question whether or not the
condensate survives the time period between the chemical
and the thermal freeze-out. This is of particular importance
if the dense matter decouples from
chemical equilibrium at temperatures $T^{ch.f.} \sim 170 - 200$ MeV
which are considerably higher than those usually associated with
the thermal freeze-out, $T^{th.f.} \sim m_\pi$. After
decoupling from chemical equilibrium, the system of stable and
long-lived particles continues to expand until it has cooled down
to the temperature $T^{th.f.}$.

For a one-dimensional scaling
expansion, the evolution of the system is determined by the equations

\ba
s \ \tau &=& {\rm const.}\label{eq:xs}\\
b \ \tau &=& {\rm const.}\label{eq:xb}\\
s_e \ \tau &=& {\rm const.}\label{eq:xcals}\\
n_\pi \ \tau  & = & {\rm const.}\label{eq:xnpi}\\
n_\alpha \ \tau  & = & {\rm const.}
\ea

which describe the conservation of entropy $S_e$, baryon number $B$,
strangeness $S$ and pion number $N_\pi$ ($s_e$, $b$, $s$
and $n_\pi$ being the corresponding densities, respectively).
The index $\alpha$ labels those of the stable particle species
($K,N,\Lambda,\omega,\eta,...$) which also have decoupled from
chemical equilibrium. For
the gas of stable particles and long-lived resonances, the
thermodynamic quantities are given by the expressions on the r.h.s.
of eqs. (\ref{eq:coupled_eq1}) -- (\ref{eq:potential}),
with $s_e = \sum_i (\varepsilon_i + P_i - \tilde{\mu}_i n_i)/T$.
Energy density and number density of the pions consist of a thermal
and a condensate component,
$\varepsilon_\pi=\varepsilon_\pi^{therm} + m_\pi n_\pi^{con}$
and $n_\pi=n_\pi^{therm} +n_\pi^{con}$, respectively.
Note that the condensate does not contribute to the entropy.
Consequently, the cooling of the system -- i.e., the function
$T(\tau)$ -- does not depend on the superfluid component.
Fig. 3 shows the cooling curves $T(\tau)$ of a pion gas, for
different values of the pion chemical potential. For comparison
we have also included the results for an ideal gas of massless
pions. It can be seen that the cooling rate increases with
increasing pion chemical potential and becomes maximal in the
presence of a condensate. This implies that Bose condensates
may reduce the lifetime of the fireball.

In particular,
we are interested in the time dependence of the
fraction of pions in the condensate,
$f_{con} \equiv n_\pi^{con}/(n_\pi^{therm}
+n_\pi^{con})$.  Eqs. (\ref{eq:xs},\ref{eq:xnpi}) imply that
for the fraction of thermal pions, $f_{therm} = 1 - f_{con}$,
\be
f_{therm}^{th.f.}  \ = \ f_{therm}^{ch.f.} \ \frac{
\left(n_\pi^{therm}/s_e)\right|_{th.f.}
}
{
\left(n_\pi^{therm}/s_e)\right|_{ch.f.}
}
\label{eq:ftherm}
\ee
where the labels $th.f.$ and $ch.f.$ refer to thermal and chemical
freeze-out as before. The condensate survives until thermal decoupling
if the r.h.s. of eq.(\ref{eq:ftherm}) remains below $1$.
For the two limiting cases
of a non-relativistic and of a ultrarelativistic gas,
$f_{con}(\tau)=$const., i.e., the fraction of particles
in the condensate does not change during the expansion.
For the pion gas,
it turns out that
in the temperature range between 200 MeV
and 150 MeV, $f_{therm}$ varies by less than $10\%$.

To summarize, we have shown that the excited hadronic matter
created in ultrarelativistic heavy ion collisions could form
a pion condensate if (i) the system is rich in baryon number
and entropy density, (ii) the hadrons decouple from local
chemical equilibrium earlier than from thermal equilibrium,
as suggested in \cite{leutwyler},  and (iii) the loss of
chemical equilibrium occurs at temperatures of $\sim 180$ MeV
or higher.  The effect
depends on the chemical decoupling
temperature, on the baryonic density and, rather sensitively,
on the resonance contribution to the
EOS.\footnote{Indeed, we have checked that for the EOS discussed in
ref.\cite{uw},
which was obtained by joining a parametrization of lattice QCD data
to a Hagedorn type resonance gas EOS, about $40\%$ of the pions
would turn out to be in the condensed phase at normal freeze-out
temperatures of $\sim 140$ MeV.}
The effect is increased if the chemical equilibrium for
strange particles is not complete, i.e. if the amount of strangeness
which is initially zero and starts to increase during the collision and
expansion process has
not yet reached its equilibrium value.

It is not the purpose of the present study to prove that
such a decoupling process occurs. To do so would require a
detailed knowledge of the density and temperature dependence
of the inelastic and elastic cross sections of hadrons in the
dense matter. Rather, we would like to point out here that,
if this decoupling happens in a certain
temperature region (like it was quoted in
\cite{leutwyler}), then a pion condensate should appear.

If a pion condensate is created in a heavy ion collision, it
may be distinguishable from the case of a resonance gas in
thermal and chemical equilibrium essentially by the
following effects.
\begin{itemize}
\item It should lead to a coherent component which then would
appear in the two-particle correlation functions of identical
pions through a
reduction of the intercept of the correlation function and the
appearance of a two exponent behaviour of the correlation function
\cite{weiner}.

\item The condensate component is moving with the fluid velocity
and has no additional thermal component which smears out the
distribution.
This might lead to characteristic bumps and
shoulders in the
rapidity and transverse momentum distributions of
pions\footnote{In ref.\cite{katrus} it was suggested that at $\mu_\pi=m_\pi$
the {\it thermal} component may lead to a dip at $p_\perp=0$ (cf. also
refs. given in \cite{katrus})}.
One also can expect that the coherent component
disappears if one considers particles of velocities that
exceed the maximum fluid velocities.

\item The lifetime of the fireball may be reduced.
This could at least partially compensate the enhancement
of the lifetime expected \cite{bertsch1}
in the case that quark-gluon-plasma
is formed  in the intial stage.
\end{itemize}

Instructive and helpful discussions with R.M. Weiner, F. Navarra and
H. Leutwyler are gratefully acknowledged. This work was supported in part
by the Deutsche Forschungsge\-mein\-schaft (DFG),
the Federal Minister for Research and Technology (BMFT) under the
contract no. 06MR731,
the Gesellschaft f\"ur Schwerionenforschung (GSI) and the Los Alamos
National Laboratory.

\newpage

{\bf FIGURE CAPTIONS}\\

{\uu{Figure 1:} The build-up of a pion chemical potential in
an expanding gas of pions and rho-mesons out of chemical equilibrium.\\

{\uu{Figure 2:}
Left column: Dependence of
the chemical potentials for baryons (dashed),
strange\-ness (dotted) and pions (solid) on the
chemical freeze-out temperature $T^{ch.f.}$, for three
different values of the baryon number density $b$. Right column:
The thermal (solid) and condensate
(dashed) components of the pion number densities at chemical
freeze-out.\\

{\uu{Figure 3:}
The temperature evolution of a pion gas for different
chemical potentials $\mu_\pi$ and condensate contributions. For
comparison, the curve for an ideal gas of massless pions has been
included as well.

\end{document}